\documentclass[9pt,letterpaper,twocolumn]{article}

\usepackage{enumerate}

\usepackage[hyphens]{url}
\usepackage{hyperref}
\usepackage[usenames,dvipsnames]{xcolor}
\hypersetup{citecolor=blue,linkcolor=blue}
\usepackage{amsmath,amsopn,amssymb}
\usepackage{subcaption}
\usepackage{endnotes,microtype,xspace,graphicx,fancyvrb,multirow}
\usepackage{epstopdf}
\usepackage{booktabs}
\usepackage{array,underscore,relsize}
\usepackage[T1]{fontenc}
\usepackage{times}
\usepackage{fancyhdr,lastpage}
\usepackage{enumitem}
\usepackage[labelfont=bf,font=small,skip=5pt]{caption}

\usepackage{fp}
\usepackage{siunitx}

\usepackage{balance}

\usepackage{geometry}
\newcommand{\cc}[1]{\mbox{\smaller[0.5]\texttt{#1}}}

\fvset{fontsize=\scriptsize,xleftmargin=8pt,numbers=left,numbersep=5pt}

\input{code/fmt}
\newcommand{\figrule}{\hrule width \hsize height .33pt}
\newcommand{\coderule}{\vspace{0em}\figrule\vspace{0.2em}}

\def\Snospace~{\S{}}

\newif\ifdraft\drafttrue
\newif\ifnotes\notestrue
\ifdraft\else\notesfalse\fi

\input{glyphtounicode}
\newcolumntype{R}[1]{>{\raggedleft\let\newline\\\arraybackslash\hspace{0pt}}p{#1}}

\newcommand{\squishlist}{
\begin{itemize}[noitemsep,nolistsep]
  \setlength{\itemsep}{-0pt}
}
\newcommand{\squishend}{
  \end{itemize}
}

\usepackage{tikz}

\newcommand*\BC[1]{%
\begin{tikzpicture}[baseline=(C.base)]
\node[draw,circle,fill=black,inner sep=0.2pt](C) {\textcolor{white}{#1}};
\end{tikzpicture}
}

\usepackage{xstring}
\newcommand{\PP}[1]{
\vspace{2px}
\noindent{\bf \IfEndWith{#1}{.}{#1}{#1.}}
}

\newcommand{\V}{\checkmark}
\newcommand{\X}{{\footnotesize $\times$}\xspace}

\newcommand{\etal}{\textit{et al}.\xspace}
\newcommand{\ie}{\textit{i}.\textit{e}.}
\newcommand{\eg}{\textit{e}.\textit{g}.}

\begin{document}

\title{Inferring Fine-grained Control Flow Inside\\ SGX Enclaves with Branch Shadowing}

\ifdefined\DRAFT
 \pagestyle{fancyplain}
 \lhead{Rev.~\therev}
 \rhead{\thedate}
 \cfoot{\thepage\ of \pageref{LastPage}}
\fi


\author{
Sangho Lee$^\dagger$\;
Ming-Wei Shih$^\dagger$\;
Prasun Gera$^\dagger$\;
Taesoo Kim$^\dagger$\;
Hyesoon Kim$^\dagger$\;
Marcus Peinado$^\ast$\;
\\\\
\emph{$^\dagger$  Georgia Institute of Technology} \\
\emph{$^\ast$ 	 Microsoft Research}\\
\\
A revised version of this paper will be\\
presented at USENIX Security Symposium 2017.\\
Please cite this paper as\\
Sangho Lee, Ming-Wei Shih, Prasun Gera, Taesoo Kim, Hyesoon Kim, and Marcus Peinado,\\
``Inferring Fine-grained Control Flow Inside SGX Enclaves with Branch Shadowing,''\\
in Proceedings of the 26th USENIX Security Symposium (Security), \\
Vancouver, Canada, August 2017.
}

\date{}
\maketitle

\begin{abstract}
  Intel Software Guard Extension (SGX)
  is a hardware-based trusted execution environment (TEE)
  that enables secure computation
  without trusting any underlying software,
  such as operating system or even hardware firmware.
  It provides strong security guarantees,
  namely, confidentiality and integrity,
  to an enclave (\ie, a program running on Intel SGX)
  through solid hardware-based isolation.
  However, a new controlled-channel attack (Xu~\etal, Oakland 2015), 
  although it is an out-of-scope attack
  according to Intel SGX's threat model,
  demonstrated that
  a malicious OS can infer
  coarse-grained control flows of an enclave
  via a series of page faults,
  and such a side-channel can be severe
  for security-sensitive applications.

  In this paper,
  we explore a new, yet critical,
  side-channel attack against Intel SGX, 
  called a branch shadowing attack, 
  which can reveal fine-grained control flows
  (\ie, each branch) of an enclave program
  running on real SGX hardware.
  The root cause of this attack is that
  Intel SGX does not clear the branch history
  when switching from enclave mode to non-enclave mode,
  leaving the fine-grained traces to the outside world through
  a branch-prediction side channel.
  %
  However, exploiting the channel
  is not so straightforward in practice because
  1) measuring branch prediction/misprediction penalties
  based on timing is too inaccurate to distinguish
  fine-grained control-flow changes and
  2) it requires sophisticated control
  over the enclave execution
  to force its execution to the interesting code blocks.
  %
  To overcome these challenges,
  we developed two novel exploitation techniques:
  1) Intel PT- and LBR-based history-inferring techniques
  and 2) APIC-based technique to control
  the execution of enclave programs in a fine-grained manner.
  %
  As a result, we could demonstrate our attack
  by breaking recent security constructs,
  including ORAM schemes, Sanctum, SGX-Shield, and T-SGX.
  %
  Not limiting our work to the attack itself,
  we thoroughly studied the feasibility
  of hardware-based solutions (\eg, branch history clearing)
  and also proposed a software-based countermeasure, called Zigzagger,
  to mitigate the branch shadowing attack in practice.

\end{abstract}

\section{Introduction}
\label{s:intro}

Establishing a trusted execution environment (TEE) is one of the most
important security requirements, as we cannot fully trust the
underlying computing platform, such as the public cloud and possibly
compromised operating system (OS).
When we want to run security-sensitive applications (\eg, processing
financial or health data) in the public cloud, we need to either fully
trust the operator, which is
problematic~\cite{cloud-security-privacy}, or encrypt all data
before uploading it to the cloud and perform computations directly on
the encrypted data by using fully homomorphic encryption, which is too
slow~\cite{popa:thesis}, or property-preserving or searchable
encryption, which is basically
weak~\cite{naveed:attack,david:searchable,grubbs:breaking}.
Even when we use the private cloud or personal workstation, similar
problems still exist because we cannot ensure whether the underlying
OS is robust against attacks due to its huge code base
and high complexity~\cite{song:kenali,jang:drk,gruss:prefetch,lu:unisan,hund:timing,kernel-self-protection}.
Since the OS is the trusted computing base (TCB), by
compromising it, an attacker can have full control of any
applications running on the platform.

Hardware-based TEEs, such as Trusted Platform Module (TPM)~\cite{tpm},
ARM TrustZone~\cite{arm:trustzone}, and Intel Software Guard Extension
(SGX)~\cite{intel:sgx-r2}, have been actively proposed to realize
TEEs.
Especially, Intel SGX is receiving a lot of attention because of its
availability and applicability.
All Intel Skylake and Kaby Lake CPUs support Intel SGX, and processes
secured by Intel SGX (\ie, processes running inside an
\emph{enclave}) can use almost every unprivileged CPU instruction
without restrictions.
As far as we can trust the hardware vendors (\ie, if there is no
hardware backdoor~\cite{yang:a2}), it is believed that the
hardware-based TEE is secure since compromising hardware is more
difficult than software in most cases due to physical limitations
(\eg, desoldering CPU packaging) and verifiability.

Unfortunately, recent
studies~\cite{xu:cca,shinde:pigeonhole} show that Intel SGX
suffers from a noise-free side-channel attack, known as a
\emph{controlled-channel attack}.
Intel SGX allows an OS to have full control of the page table of
an SGX program; it can map or unmap arbitrary memory pages of the SGX
program.
This makes a malicious OS know exactly which memory pages a victim SGX
program attempts to access by monitoring page faults.
Unlike conventional side channels such as cache-timing channels,
this page-fault side channel is deterministic; namely, it does not
suffer from any measurement noise.

The controlled-channel attack has a limitation; it only reveals
coarse-grained \emph{page-level access patterns}.
Intel said that its architecture (including Intel SGX) aims to
provide protection against side-channel attacks at the cache-line
granularity~\cite{intel-side-channel}.
Thus, such page-level access information would be too coarse grained
to be their main concern.
Further, researchers  propose effective countermeasures against
the controlled-channel attack,
which are based on balanced execution~\cite{shinde:pigeonhole}
and user-space page-fault
detection~\cite{tsgx,costan:sanctum,shinde:pigeonhole}.
However, these countermeasures only prevent the
controlled-channel attack, hence a fine-grained side-channel attack, if
it exists, would easily bypass them.

We thoroughly explored Intel SGX to know whether it has
a critical side channel that reveals fine-grained information (finer
than cache-line granularity) and is robust against noise.
One key observation is that Intel SGX \emph{leaves branch history uncleared
during enclave mode switches}, which can be used as a side channel.
Knowing the branch history (\ie, taken or not-taken branches) is
critical because it would reveal the fine-grained execution trace of a
process in terms of basic blocks.
To avoid this problem, Intel SGX hides the branch history information
inside an enclave from hardware performance counters, including last
branch record (LBR) and Intel Processor Trace
(PT)~\cite{intel:sgx-r2}.
In other words, an OS is unable to directly monitor and manipulate the
branch history of enclave processes.
However, since Intel SGX does not clear the branch history, the
fine-grained execution traces can be potentially inferred outside of
an enclave through a branch-prediction side
channel~\cite{aciicmez07:branch,evtyushkin:branch-covert,evtyushkin:btb}.

The branch-prediction side channel attack aims to recognize whether the
history of a target branch instruction is stored into a CPU internal
buffer for the branch prediction, known as the branch target buffer
(BTB).
To achieve the goal, this attack measures how long it takes to execute
a shadowed branch instruction, which could be mapped into the same BTB
entry the target branch instruction is stored into due to their same
address in terms of lowest 31 bits (\autoref{subs:branch-prediction}) or
set conflicts (\autoref{subs:advanced}).
This collision between two branch instructions results in
a timing difference due to \emph{branch misprediction penalty}
(\autoref{s:attack}).
Several researchers have tried to use this side channel
to infer cryptographic keys~\cite{aciicmez07:branch}, create a covert
channel~\cite{evtyushkin:branch-covert}, and break address space
layout randomization (ASLR)~\cite{evtyushkin:btb}.

This attack, however, is difficult to realize without a compromised OS
(\ie, the threat model of SGX) and a precise measurement strategy due
to the following reasons.
First, an attacker cannot easily guess the address of a target branch
instruction and manipulates its branch addresses due to ASLR.
Second, since the BTB's capacity is limited, its entry would be easily
overwritten by other branch instructions before an attacker gets a
chance to probe it.
Third, 
measuring branch misprediction penalty suffers from tremendous time
noise (\autoref{subs:conditional}).
In summary, an attacker should have 1) a right to freely manipulate
the virtual address space, 2) access to the BTB anytime before it is
overwritten, and 3) a method to recognize branch misprediction with
negligible (or no) noise.

In this paper, we present a new branch-prediction side-channel attack,
called the \emph{branch shadowing attack}, to identify fine-grained control
flows inside an enclave without noise (to identify conditional and
indirect branches) or with negligible noise (to identify unconditional
branch).
A malicious OS can easily manipulate the virtual address space of an
enclave process, so that it is easy to create shadowed branch
instructions colliding with target branch instructions in an enclave.
To minimize the measurement noise, we tried to use Intel PT's
timestamps instead of \cc{RDTSC} (\autoref{subs:conditional}).
More importantly, we found that Skylake's LBR allows us to obtain the most
accurate information for the branch shadowing attack because it
reports whether each conditional/indirect branch instruction is
correctly predicted or mispredicted.
That is, we can \emph{exactly} know the prediction and misprediction of
conditional and indirect branches (\autoref{subs:conditional},
\autoref{subs:indi-shadow}).
Furthermore, Skylake's LBR reports elapsed core cycles between LBR
entry updates, which are very stable according to our measurements
(\autoref{subs:conditional}).
By using this information, we can precisely infer the execution of an
unconditional branch (\autoref{subs:uncondi-shadow}).

Precise execution control and frequent branch history probing are
other important requirements of the branch shadowing attack.
To achieve these goals, we manipulated the frequency of the local
advanced programmable interrupt controller (APIC) timer as frequently as
possible and modified the timer interrupt code to make it execute the
branch shadowing attack.
Further, we selectively disable the CPU cache when a more precise
attack is needed (\autoref{subs:timer}).

We performed case studies to evaluate the effectiveness of the branch
shadowing attack (\autoref{s:study}).
First, we extracted sensitive information from
SGX applications including Linux SGX
SDK (string conversions and formatted strings),
mbed TLS cryptographic library (RSA private keys),
LIBSVM machine-learning library (classification models and parameters),
and Apache web server (HTTP requests).
Next, we analyzed state-of-the-art studies to secure SGX including
deterministic multiplexing~\cite{shinde:pigeonhole},
Sanctum~\cite{costan:sanctum}, SGX-Shield~\cite{sgx-shield}, and
T-SGX~\cite{tsgx}, and confirmed that our attack bypassed all of
them.
Finally, we suggested hardware- and software-based countermeasures
against the branch shadowing attack, by
clearing branch history during enclave mode switches and using
indirect branches with multiple targets (\autoref{s:counter}).
Both of them had acceptable overhead (below 1.3\X).

In summary, the contributions of this paper are as follows:

\squishlist
\item \textbf{Fine-grained attack.}
  We demonstrate that the branch shadowing attack can identify
  fine-grained control flow information inside an enclave in terms of
  basic blocks, unlike the state-of-the-art controlled-channel attack
  that only reveals page-level accesses.
\item \textbf{Precise attack.}
  We make the branch shadowing attack very precise by 1) exploiting
  Intel PT and LBR to correctly identify branch history and 2)
  adjusting the local APIC timer to precisely control the execution
  inside an enclave.
  We can deterministically know whether a target branch has taken or
  not taken without noise (conditional and indirect branches)
  or with negligible noise (unconditional branch).
\item \textbf{Countermeasures.}
  We design proof-of-concept hardware- and software-based
  countermeasures against the branch shadowing attack.
  We evaluate both approaches' effectiveness and performance overhead.
  \squishend

The remainder of this paper is organized as follows.
\autoref{s:background} explains details about Intel SGX and other
processor features our attack relies on.
\autoref{s:attack} introduces our branch shadowing attack in detail.
\autoref{s:study} explains how the branch shadow attack reveals
sensitive information from SGX applications and defeats recent security
proposals.
\autoref{s:counter} describes our hardware-based and software-based
mitigations against the branch shadowing attack.
\autoref{s:discuss} discusses some limitations of the branch shadowing
attack and considers possible advanced attacks.
\autoref{s:relwk} introduces related work.
\autoref{s:conclusion} concludes this paper.

\section{Background}
\label{s:background}
In this section, we first explain the basics of Intel SGX.
Then, we explain Intel CPU's other essential features (branch
prediction, LBR, and local APIC timer) related with our attack.

\subsection{Intel SGX}
Intel SGX~\cite{costan:sgx-explained} is one of the existing
implementations of hardware-based TEE that has been shipped with Intel
CPU since Skylake.
SGX is designed under the assumption that the TCB is reduced to include only the internals
of the CPU package, \ie, privileged software such as OS or hypervisor
and other hardware units are excluded.
To this end, SGX allows an application to instantiate a secure container
called an \emph{enclave}.
Enclaves are allocated in a dedicated physical memory region, called the
\emph{enclave page cache} (EPC), that is protected by an on-chip memory
encryption engine (MEE) such that the EPC content always stays encrypted
and is only decrypted right before entering the CPU package.
SGX also enforces different CPU access controls between enclave code and
non-enclave code to allow only the enclave to access its own code and data,
while accesses from other software are prohibited. Note that the enclave
is still allowed to access non-enclave memory region.
Enclaves can be created as part of applications' address space via an SGX
instruction set. Measurement of code and data is calculated
during the loading process and can serve as evidence
about the enclave in remote attestation~\cite{intel-attestation, intel-attestation-api}.

\PP{Non-enclave cod and enclave code interaction.}
Non-enclave code can only switch to enclave code via either
the \cc{EENTER} instruction to a list of defined entry points
or the \cc{ERESUME} instruction that resumes execution
where an \emph{asynchronous enclave exit} (AEX) happens due
to events such as exceptions and interrupts.
Upon enclave exit (AEX or using the \cc{EEXIT} instruction),
a series of checks and actions is performed, such as TLB flush,
to ensure the isolation of an enclave.
To exchange input and output values,
enclave code and non-enclave code use untrusted memory outside an enclave.


\subsection{Branch Prediction}
\label{subs:branch-prediction}
Branch prediction is one of the most important features of modern
pipelined processors.
Basically, an instruction pipeline consists of four major stages:
fetch, decode, execute, and write-back.
This pipeline structure makes the processor execute an instruction
while fetching/decoding the next instructions and storing the result of
the previous instruction into the memory (or the cache); namely, the
processor can execute a number of instructions in parallel.
However, the pipelined processor has a problem with a branch
instruction because, before executing it, the processor cannot know
what the next instruction is.
Making the instruction pipeline stall until the processor confirms the
next instruction is bad for the overall throughput, so modern
processors have a branch prediction unit (BPU) to
\emph{predict} the next instruction after a branch
instruction and execute it to maintain the pipeline utilization.
However, a branch misprediction would bring a penalty because the
processor needs to clear the pipeline and roll back the execution
results.
This is why the Intel optimization manual~\cite{intel-optimization}
emphasizes branches and Intel provides a dedicated hardware feature to
log branch information: the LBR, which will be
explained later.

\PP{Branch and branch target prediction}
There are two kinds of branch predictions: \emph{branch prediction}
and \emph{branch target prediction}.
Branch prediction is a procedure to predict the next instruction
of a conditional branch by guessing whether it will be taken or not be
taken.
Branch target prediction is a procedure to predict the target
instruction of a conditional or unconditional branch before executing
it.
For branch target prediction, modern processors have the branch
target buffer (BTB) to store the computed target addresses of taken
branch instructions and fetch them when the corresponding branch
instructions are about to be executed.


\PP{BTB structure and partial tag hit}
The BTB resembles as cache. Some address bits are
used to compute the index bits and some address bits are used for
tag. However, in the BTB, only smaller number of bits are used for tag
to save the BTB unlike cache uses the all the
remaining bits for tag. For example, in 64 bit address space, if
ADDR[11:0] are used for index, instead of using ADDR[63:12] for a tag,
only partial number of bits such as ADDR[31:12] are used as tag. The
reasons are first, compared to a data cache, the BTB size is very
small, which results in many unused bits. Second, typically in one
program, the upper bits are almost the same. Third, unlike a cache
which needs to provide the architectural values, the BTB
is just a predictor. Even if a partial tag matching results in a false
BTB hit, the correct target will be computed at the execution stage
and the pipeline will be roll back if the prediction is wrong. This
feature is needed because for indirect branches, even a BTB hit can
results in a wrong prediction, which should be corrected at the
execution stage.

\PP{Static and dynamic branch prediction}
The static branch prediction is a basic rule of predicting the next
instruction of a branch instruction when it has no
history~\cite{intel-optimization}.
First, a processor predicts a conditional branch will not
be taken, which means the next instruction will be directly
fetched (\ie, a fall-through path). 
Second, a processor predicts an indirect branch will not
be taken.
Third, a processor predicts that an unconditional branch will
be taken (\ie, the specified target will be fetched).

When a branch instruction has a history, \ie, it has a BTB entry, a
processor predicts that the stored target address will be the next
instruction.
This procedure is known as dynamic branch prediction.
Note that the changes of branch prediction behaviors according to the
branch history can be used as a side channel to infer a victim
process's activities (\autoref{s:attack}).

\subsection{Last Branch Record (LBR)}
\label{s:lbr}
The LBR is Intel CPU's new feature that logs the information of
recently \emph{taken} branches without any performance degradation, as
it is separated from the instruction
pipeline~\cite{lbr,lbr2,intel-doc}.
In Skylake CPUs, the LBR stores the information of up to 32 recent
branches, including the address of a branch instruction (from), the
target address (to), whether the branch or branch target was
mispredicted, and the elapsed core cycles between LBR entry updates
(also known as the timed LBR).
Without filtering, the LBR records all kinds of branches, including
function calls, function returns, indirect branches, and conditional
branches.
Also, the LBR can selectively record branches taken in the user space,
kernel space, or both.
%


\subsection{Local APIC Timer}
\label{s:lapic}
The local advanced programmable interrupt controller (APIC) is a
component of Intel CPUs to configure and handle CPU-specific
interrupts~\cite[\S10]{intel-doc}.
An OS can program the local APIC through memory-mapped registers
(\eg, device configuration register) or model-specific registers
(MSRs) to adjust the frequency of the \emph{local APIC timer}, which
generates high-resolution timer interrupts, and deliver an interrupt
to a CPU core (\eg, inter-processor interrupt (IPI) and I/O interrupt
from the I/O APIC).

Intel CPUs support three local APIC timer modes: periodic, one-shot,
and timestamp counter (TSC)-deadline modes.
The periodic mode lets an OS configure the initial-count register
whose value is copied into the current-count register the local APIC
timer uses.
The current-count register's value decreases at the rate of the bus
frequency, and when it becomes zero, a timer interrupt is generated and
the register is re-initialized by using the initial-count register.
The one-shot mode lets an OS reconfigure the initial-count counter
value whenever a timer interrupt is generated.
The TSC-deadline mode is the most advanced and precise timer mode that
allows an OS to specify when the next timer interrupt should occur in
terms of a TSC value.
Our target Linux system (kernel version 4.4) uses the TSC-deadline
mode, so we mainly considers this mode.

\section{Branch Shadowing Attacks}
\label{s:attack}
In this section, we explain our attack, the branch shadowing attack,
to obtain the fine-grained control flow information of an enclave process.
We first introduce our threat model and depict how we can attack three
types of branches: conditional, unconditional, and indirect branches.
Then, we describe our approach to synchronize the victim and attack
code in terms of execution time and memory address space.

\subsection{Threat Model}
\label{subs:model}
We explain our threat model, which is based on the original threat
model of Intel SGX and the controlled-channel attack~\cite{xu:cca}: an
attacker has compromised the operating system and exploits it to
attack a target enclave program.

First, the attacker knows the possible control flows of a target
enclave program (\ie, a sequence of branch instructions and their
targets) by statically or dynamically analyzing its source code or
binary.
Unobservable code (\eg, self-modifying code and code from remote
servers) is outside the scope of our attack.
Also, the attacker can map the target enclave program into specific
memory addresses to designate the locations of each branch instruction
and its target address.
Self-paging~\cite{hand:self-paging} and live re-randomization of
address-space layout~\cite{giuffrida:os-asr} inside an enclave are outside
the scope of our attack.

Second, the attacker can infer which portion of code the target
enclave program runs through observable events,
such as calling functions outside an enclave and page faults.
Our attack uses this information to synchronize the execution of the
target enclave program with the branch probing code
(\autoref{subs:synchronization}).

Third, the attacker can interrupt the execution of the target enclave
program as frequently as possible to frequently run the branch probing
code.
This can be done by manipulating a local APIC timer and/or
disabling the CPU cache (\autoref{subs:timer}).

Fourth, the attacker can recognize the branch probing code's branch
prediction and misprediction by monitoring hardware performance
counters (\eg, the LBR) or measuring branch misprediction
penalty~\cite{aciicmez07:branch,evtyushkin:btb,evtyushkin:branch-covert}.

\subsection{Overview}
\label{subs:overview}
The goal of the branch shadowing attack is to obtain the
fine-grained control flow of an enclave program by 1) knowing whether a branch
instruction has been taken or not taken and 2) inferring the target
address of the taken branch.
To achieve the goal, an attacker first needs to analyze the source
code and/or binary of a victim enclave program to enumerate all
branches (unconditional, conditional, and indirect branches) and their
target addresses.
Next, the attacker writes shadow codes for each branch or a set of
branches to probe their branch history, which is similar to
Evtyushkin~\etal's attack using the BTB~\cite{evtyushkin:btb}.
Since using BTB alone suffers from significant noise, the branch
shadowing uses both BTB and LBR,
which allows the attacker precisely identify the states of all branch types
(\autoref{subs:conditional}, \autoref{subs:uncondi-shadow},
\autoref{subs:indi-shadow}).
Due the size limitation of BTB and LBR, the branch shadowing attack
has to synchronize the execution of the victim code and the shadow
code in terms of execution time and memory address space.
We manipulate the local APIC timer and the CPU cache
(\autoref{subs:timer}) to frequently interrupt an enclave process's
execution for synchronization, and adjust virtual address space
(\autoref{subs:memory}) and run probing code to find a function
an enclave process is running
(\autoref{subs:synchronization}).

\subsection{Conditional Branch Shadowing}
\label{subs:conditional}
\begin{figure}[t!]
  \centering
  \begin{subfigure}[t]{.48\columnwidth}
    \centering
    \input{code/condi-target.c.tex}
    \coderule
    \caption{Victim \cc{if}-\cc{else} statement executed inside
      an enclave. According to the value of \cc{a}, either \cc{if}-block
      or \cc{else}-block is executed.}
    \label{c:condi-target}
  \end{subfigure}
  ~
  \begin{subfigure}[t]{.48\columnwidth}
    \input{code/condi-shadow.c.tex}
    \coderule
    \caption{Shadowed \cc{if}-\cc{else} statement aligned with
      \textbf{(a)} and executed after \textbf{(a)}. The BPU predicts
      which block will be executed according to the branch history of
      \textbf{(a)}.}
    \label{c:condi-shadow}
  \end{subfigure}
  \caption{
    An example of a shadowing scheme (b)
    against a victim's conditional branch (a).
    The execution time
    (\ie, running [1, 5-10], marked with $\star$ in (b))
    of the shadowing instance
    depends on the branching result
    (\ie, taken or not at [1] in (a))
    of the victim instance.
  }
  \label{c:condi}
\end{figure}
\begin{figure*}[!t]
  \centering
  \begin{subfigure}[t]{\textwidth}
    \centering
    \includegraphics[width=.6\textwidth]{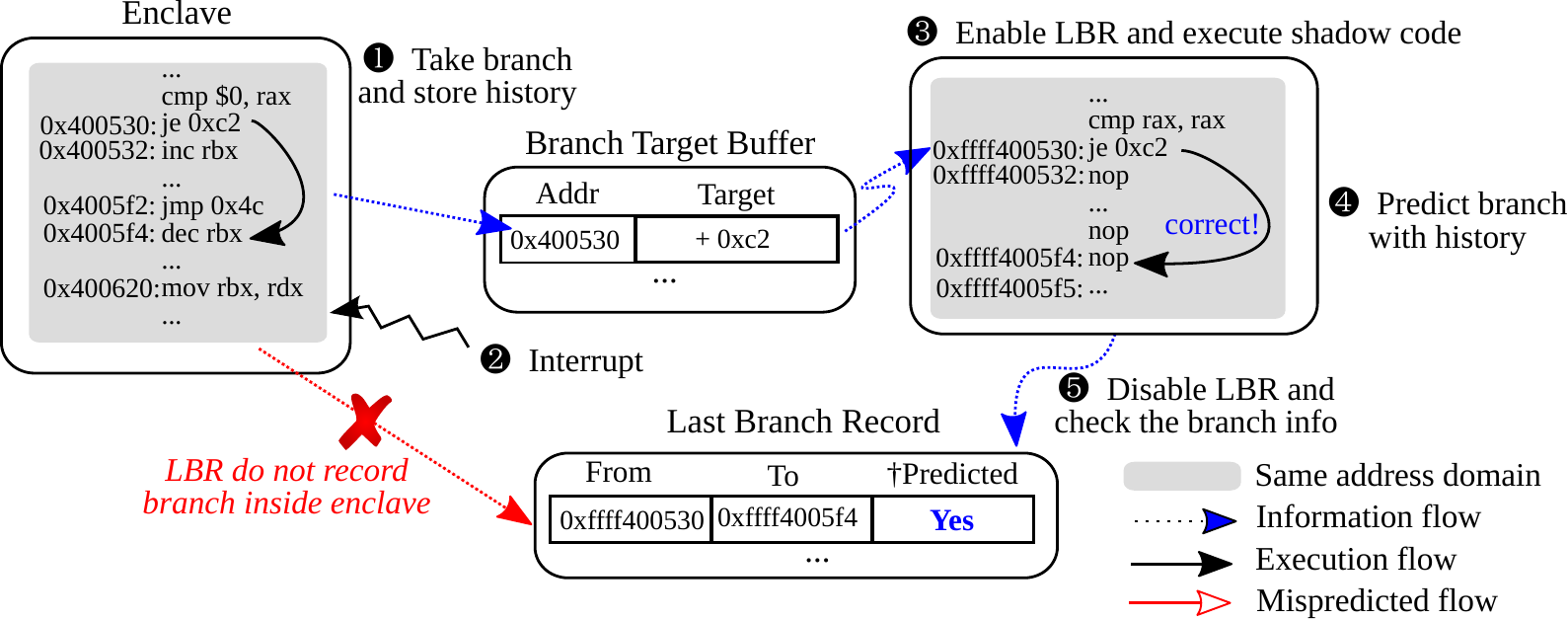}
    \caption{Case 1: The target conditional branch has been taken.}
    \label{f:condi-taken}
  \end{subfigure}
  \begin{subfigure}[t]{\textwidth}
    \centering
    \includegraphics[width=.6\textwidth]{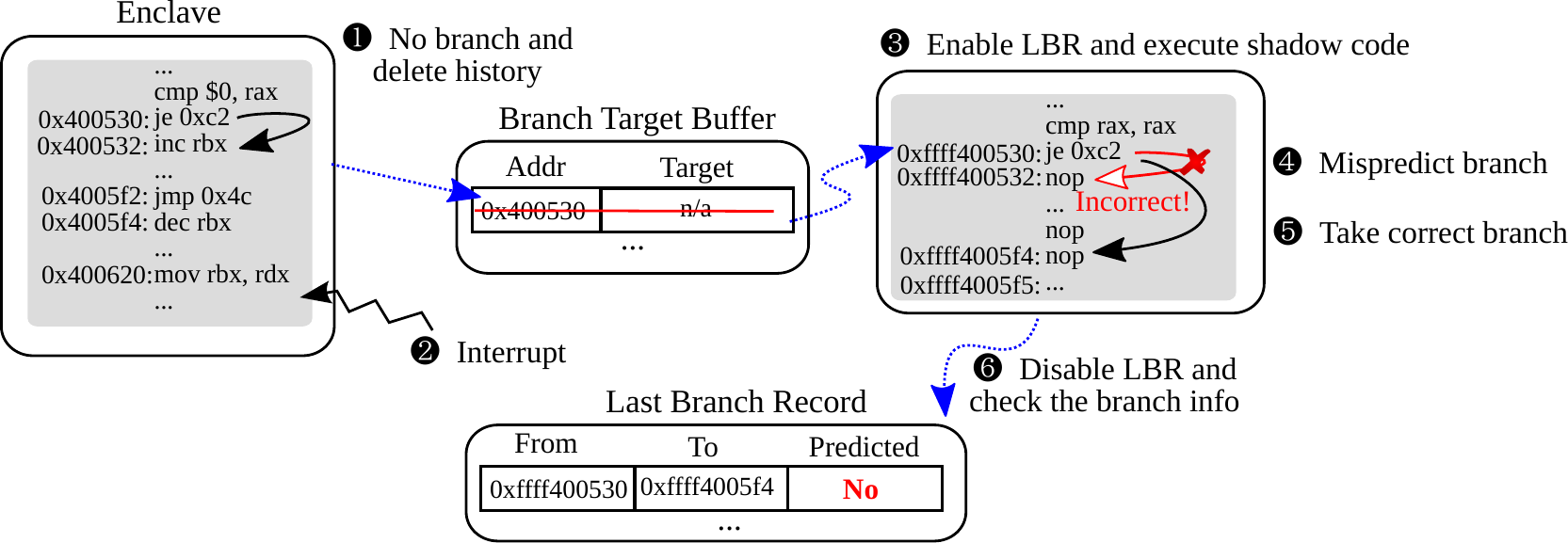}
    \caption{Case 2: The target conditional branch has not been taken
      (\ie, it has either not been executed or been executed but not
      taken.)}
    \label{f:condi-not-taken}
  \end{subfigure}
  \caption{Branch shadowing attack against a \emph{conditional branch}
    (\ie, Case 1 for taken and Case 2 for non-taken branches) inside
    an enclave ($\dagger$ LBR records the result of
    \emph{misprediction}. For clarity, we use the result of
    \emph{prediction} in this paper.)}
  \label{f:condi}
\end{figure*}
We explain how an attacker can know whether a target conditional
branch inside an enclave has been taken or not taken by shadowing its
branch history.
Unlike other branch types (unconditional and indirect branches)
explained later, a conditional branch is related to two kinds of
prediction: branch prediction and branch target prediction.
For a conditional branch, we focus on recognizing whether the branch
prediction is correct or not because it lets us know the result of the
condition evaluation (\ie, a given condition of \cc{if} statement or
\cc{for} loop).
This goal differs from the previous branch timing attack against
ASLR~\cite{evtyushkin:btb} because its goal is finding a randomized
target address of a branch instruction by probing possible target
addresses while monitoring the penalty of branch target
mispredictions.

\PP{Inferring through timing (RDTSC)}
We first explain how we can infer branch mispredictions with \cc{RDTSC},
which is based on Evtyushkin~\etal's approach~\cite{evtyushkin:btb}.
\autoref{c:condi} shows example code with a conditional branch and its
shadow for attack.
The victim code's execution depends on the value of \cc{a}: if \cc{a}
is not zero, the branch will not be taken such that the \cc{if}-block
will be executed; otherwise, the branch will be taken such that the
\cc{else}-block will be executed.
In contrast, we make the shadow code's branch always be taken (\ie,
the \cc{else}-block is always executed).
Without the branch history, this branch is always mispredicted due to
the static branch prediction rule (\autoref{subs:branch-prediction}).

To exploit the branch history, we have to align the shadow code's
address (both the branch instruction and its target address)
with the victim code's address in terms of lower 31 bits, such
that the shadow code can share the same BTB entries with the victim code.
%

When the victim code has been executed before the (aligned) shadow
code is executed, the branch prediction or misprediction of the shadow
code depends on the execution of the victim code.
If the conditional branch of the victim code has been taken, \ie, if
\cc{a} was zero, the BPU predicts that the shadow code will also take the
conditional branch, which is a correct prediction so that no rollback
will happen.
If the conditional branch of the victim code either has not been
taken, \ie, if \cc{a} was not zero, or has not been executed, the BPU
predicts that the shadow code will not take the conditional branch.
However, this is an incorrect prediction such that a rollback will
happen.

Previous branch timing attacks try to measure such a rollback penalty
by using \cc{RDTSC} or \cc{RDTSCP} instructions (\eg, before Line 1
and after Line 5 of \autoref{c:condi-shadow}).
However, according to our experiments (\autoref{tbl:timing}), branch
misprediction penalties were very noisy such that it was difficult to
set a clear boundary between correct prediction and misprediction.
This is because the number of instructions that would be mistakenly
executed due to the branch misprediction is difficult to predict
given the highly complicated internal structure of the latest Intel
CPUs (\eg, out-of-order execution).
%
%
Therefore, we think that the \cc{RDTSC}-based inference is difficult
to use in practice and, thus, we aim to use the LBR to realize precise
attacks, since it lets us know branch misprediction information and
its elapsed cycle feature has a small noise (\autoref{tbl:timing}).

\begin{table}[!t]
  \centering
  \scriptsize
  \begin{tabular}{@{}lrrrr@{}}
  \toprule
  & \multicolumn{2}{c}{\textbf{Correct prediction}} & \multicolumn{2}{c}{\textbf{Misprediction}} \\
  \cmidrule(r){2-3} \cmidrule(l){4-5}
  & \textbf{Mean} & $\sigma$      & \textbf{Mean} & $\sigma$ \\
  \midrule
  \textbf{RDTSCP}               & 94.21 & 13.10  & 120.61 & \textbf{806.56} \\
  \textbf{Intel PT CYC packets} & 59.59 & 14.44  & 90.64 &  \textbf{191.48} \\
  \textbf{LBR elapsed cycle}    & 25.69 &   9.72 & 35.04 &   \textbf{10.52} \\
  \bottomrule
\end{tabular}

  \caption{Measuring branch misprediction penalty with \cc{RDTSCP},
    Intel PT CYC packet, and LBR elapsed cycle (10,000 times). Our
    machine has an Intel Core i7 6700K CPU (4GHz). We put 120 \cc{NOP}
    instructions at the fall-through path. The LBR elapsed cycle is
    less noisy than \cc{RDTSCP} and Intel PT. $\sigma$ stands for standard deviation.}
  \label{tbl:timing}
\end{table}

\PP{Inferring from execution traces (Intel PT)}
%
%
In addition to \cc{RDTSC},
we found that Intel PT
can be used to measure 
a misprediction penalty of a target branch,
as it provides precise elapsed cycles
(known as a CYC packet)
between each PT packet.
However, this CYC packets cannot be immediately used for our purpose
because Intel PT aggregates a series of
conditional and unconditional branches
into a single packet as an optimization.
To avoid this problem,
we intentionally insert an indirect branch
right after the target branch,
making all branches
properly record their elapsed time
in different CYC packets.
As shown in \autoref{tbl:timing}, 
Intel PT's timing information about
the branch misprediction
can significantly reduce
the measurement variance of our attack
compared to \cc{RDTSCP}.

\PP{Precise leakages (LBR)}
\autoref{f:condi} shows a detailed procedure of conditional branch
shadowing with the BTB and LBR.
We first explain the case where a conditional branch has been taken
(Case 1).
\BC{1} A conditional branch of the victim code inside an enclave is
taken and the corresponding information (the branch instruction's
address and the relative target address) is stored into the BTB.
Note that this branch taken happens inside an enclave such that the LBR
does not report this information unless we run an enclave process with
a debug mode.
\BC{2} The enclave execution is interrupted and an OS takes control.
We explain how a malicious OS can frequently interrupt an enclave
process in \autoref{subs:timer}.
\BC{3} The OS kernel enables the LBR and then executes the shadow
code.
\BC{4} The BPU correctly predicts that the conditional branch will
be taken.
%
%
\BC{5} Finally, by disabling and retrieving the LBR, we can know the
shadow code's conditional branch has been \emph{correctly predicted}.
Note that, by default, the LBR reports that all the branches (including
function calls) occurred in user and kernel spaces.
Since our shadow code have no function calls and is executed in the
kernel, we use the LBR's filtering mechanism to ignore every function
call and all branches in the user space.

Next, we explain the case where a conditional branch has not been
taken (Case 2).
\BC{1} The conditional branch of the victim code inside an enclave is
not taken, so either no information is stored into the BTB or the
corresponding old information is deleted (\ie, old information can be
evicted if newer branches need to be inserted in the same set.)
\BC{2} The enclave execution is interrupted and an OS takes control.
\BC{3} The OS kernel enables the LBR and then executes the shadow code.
\BC{4} The BPU incorrectly predicts that the conditional branch
will not been taken.
\BC{5} The execution is rolled back and the code takes the branch.
%
%
\BC{6} Finally, by disabling and retrieving the LBR, we can know that
the shadow code's conditional branch has been \emph{mispredicted} and can
see the misprediction penalty.

\subsection{Unconditional Branch Shadowing}
\label{subs:uncondi-shadow}

\begin{figure*}[!t]
  \centering
  \begin{subfigure}[t]{\textwidth}
    \centering
    \includegraphics[width=.6\textwidth]{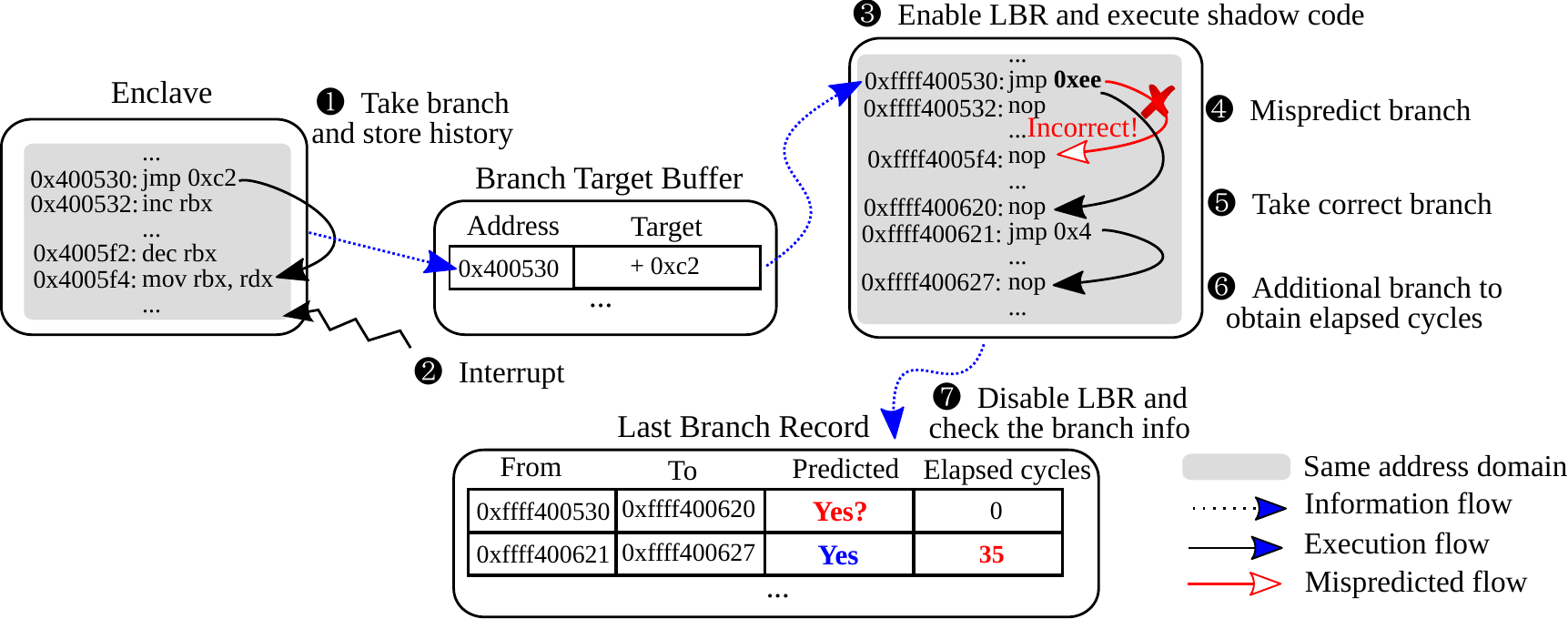}
    \caption{Case 3: The target unconditional branch has been taken.
      The LBR does not report the misprediction of unconditional
      branches, but we can infer it by using the elapsed cycles.}
    \label{f:uncondi-taken}
  \end{subfigure}
  \begin{subfigure}[t]{\textwidth}
    \centering
    \includegraphics[width=.6\textwidth]{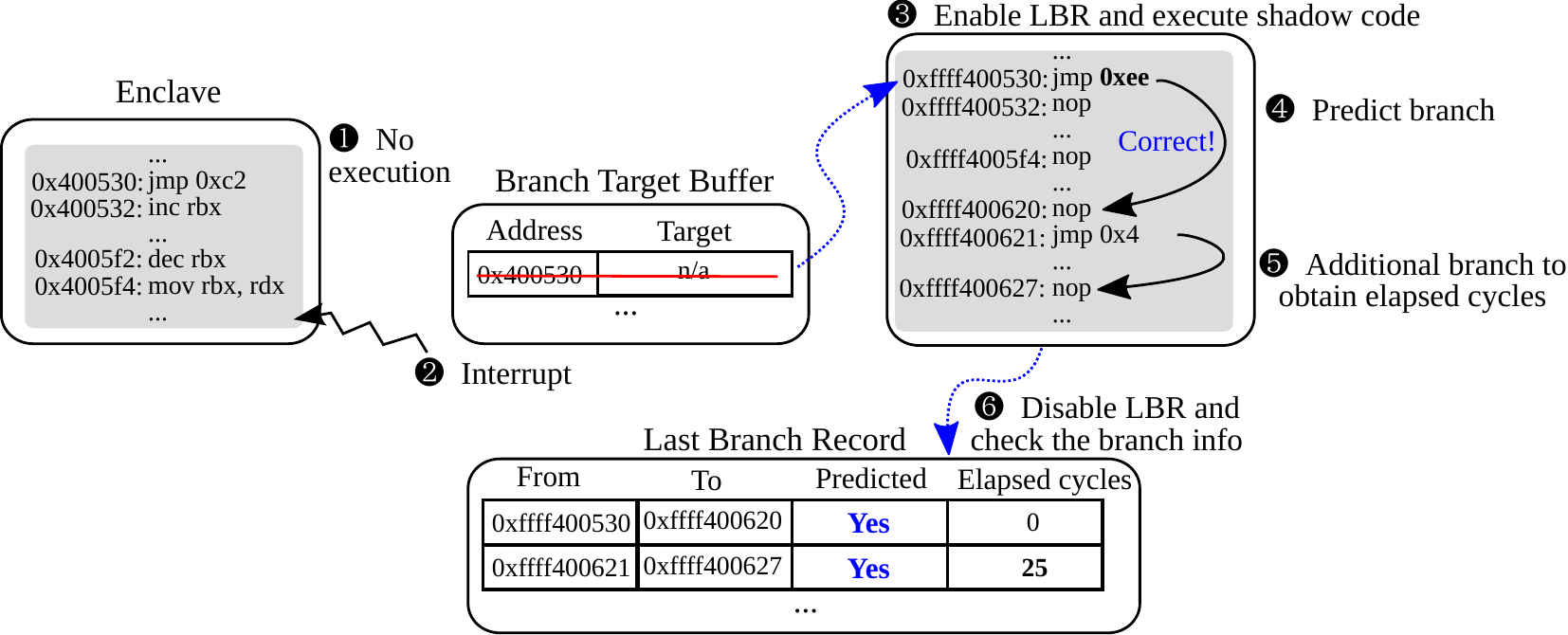}
    \caption{Case 4: The target unconditional branch has not been
      taken (\ie, it has not been executed.)}
    \label{f:uncondi-not-taken}
  \end{subfigure}
  \caption{Branch prediction attack against an unconditional branch
    inside an enclave.}
  \label{f:uncondi}
\end{figure*}

We explain how an attacker can know whether a target unconditional
branch inside an enclave has been executed or not by shadowing its
branch history.
The execution of an unconditional branch gives us two
kinds of information.
First, an attacker can infer where the instruction pointer (IP) inside
an enclave currently points.
Second, an attacker can infer the result of the condition evaluation
of an \cc{if-else} statement because an \cc{if} block's last
instruction is an unconditional branch to skip the corresponding
\cc{else} block.

Unlike a conditional branch, an unconditional branch is always taken
when it is executed; \ie, a branch prediction is not needed.
Thus, to recognize its behavior, we need to divert its target address
to observe branch target mispredictions, not branch mispredictions.
Interestingly, we found that the LBR does not report the branch target
misprediction of an unconditional branch, unlike conditional and
indirect branches.
Thus, we use the elapsed cycles of a branch that the LBR reports to
identify branch target misprediction penalty, which is less noisy than
\cc{RDTSC} (\autoref{tbl:timing}).

\PP{Attack procedure}
\autoref{f:uncondi} shows a procedure of unconditional branch
shadowing.
Unlike the conditional branch shadowing, we make the target of the
shadowed unconditional branch differ from the target of the victim
unconditional branch inside an enclave to monitor a branch target
misprediction of the shadowed branch.
We first explain the case where an unconditional branch has been
executed (Case 3).
\BC{1} An unconditional branch of the victim code inside an enclave
is executed (\ie, taken) and the corresponding information is stored
into the BTB.
\BC{2} The enclave execution is interrupted and OS takes control.
\BC{3} The OS kernel enables the LBR and then executes the shadow code.
\BC{4} The BPU mispredicts the branch target of the shadowed
unconditional branch due to the mismatched branch history.
\BC{5} The execution is rolled back and the shadow code jumps into the
correct target.
\BC{6} The shadow code executes an additional branch to measure the
elapsed cycle of the mispredicted branch.
\BC{7} Finally, by disabling and retrieving the LBR, we infer that
a branch target misprediction happened, according to the large elapsed
cycles.

Next, we explain the case where an unconditional branch has not been
taken (Case 4).
\BC{1} The enclave does not execute an unconditional branch of the
victim code (\eg, a function containing the code is never executed),
so the BTB does not have any information of the branch.
\BC{2} The enclave execution is interrupted and an OS takes control.
\BC{3} The OS kernel enables the LBR and then executes the shadow code.
\BC{4} The BPU correctly predicts the unconditional branch's
target because no branch history exists.
\BC{5} The shadow code executes an additional branch to measure the
elapsed cycles.
\BC{6} By disabling and retrieving the LBR, we infer that
no branch target misprediction happened, according to the small elapsed
cycles.%

\PP{No misprediction of unconditional branch}
We found that the LBR always reports that every taken
unconditional branch has been correctly predicted no matter whether a
branch target misprediction has happened due to the BTB collision.
Intel does not explain about this behavior, but we think
that this is because the target of an unconditional branch is fixed such
that it should not be mispredicted in general.
The LBR is proposed for profiling branches and letting programmers
know which branches are frequently mispredicted.
With this information, they can improve the performance of
their program by reorganizing branches to reduce mispredictions.
In contrast, programmers have no way to handle mispredicted
unconditional branches, which depend on the execution of kernel or
another process simultaneously running in the same core due to
hyperthreading; \ie, it does not help programmers improve their
program and only reveals side-channel information.
We believe these are the reasons the LBR just treats every unconditional
branch as correctly predicted.


\subsection{Indirect Branch Shadowing}
\label{subs:indi-shadow}
\begin{figure*}[!t]
  \centering
  \begin{subfigure}[t]{\textwidth}
    \centering
    \includegraphics[width=.6\textwidth]{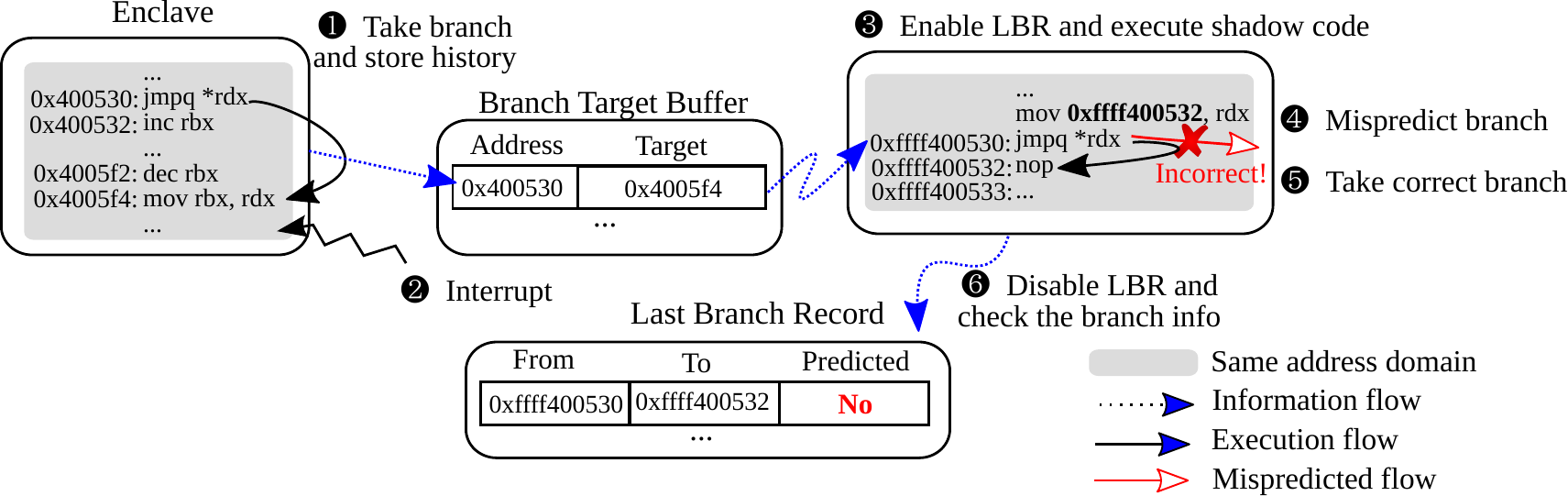}
    \caption{Case 5: The target indirect branch has been taken.}
    \label{f:indi-taken}
  \end{subfigure}
  \begin{subfigure}[t]{\textwidth}
    \centering
    \includegraphics[width=.6\textwidth]{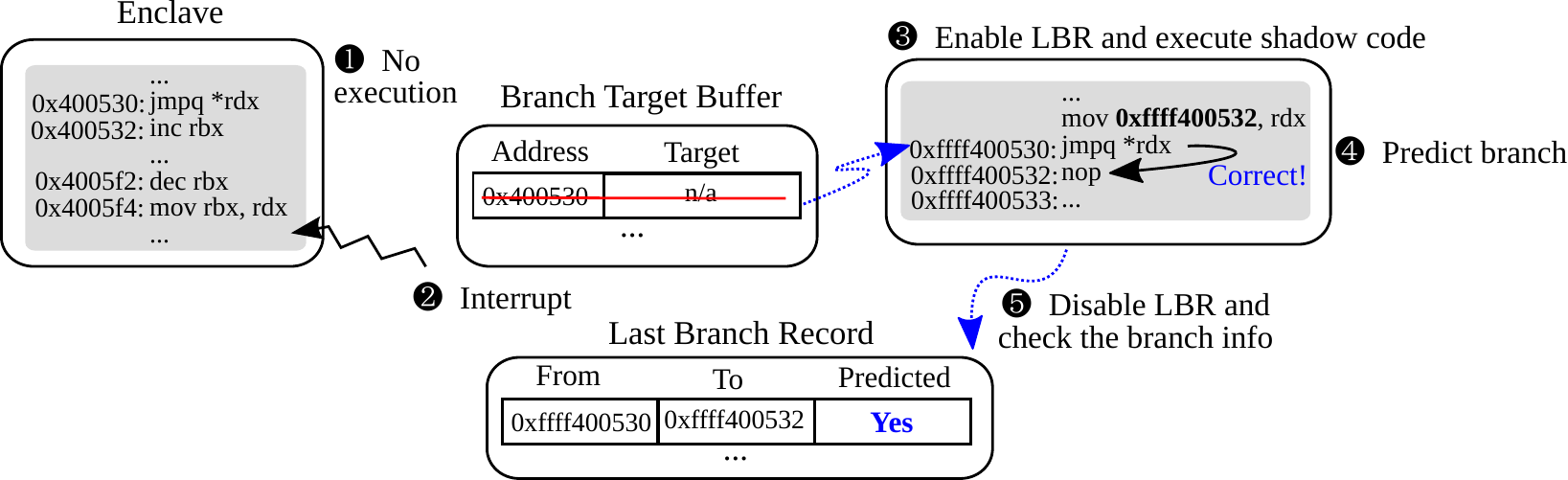}
    \caption{Case 6: The target indirect branch has not been
      taken (\ie, it has not been executed.)}
    \label{f:indi-not-taken}
  \end{subfigure}
  \caption{Branch prediction attack against an indirect branch
    inside an enclave}
  \label{f:indi}
\end{figure*}

We explain how we can know whether a target indirect branch inside an
enclave has been executed by shadowing its branch history.
Like an unconditional branch, an indirect branch is always taken when
it is executed.
However, unlike an unconditional branch, an indirect branch has no
fixed branch target.
If there is no history the BPU predicts that the right next
instruction will be executed; this is the same as the indirect
branch not being taken.
To recognize its behavior, we make a shadowed indirect
branch target its next instruction to monitor a branch target
misprediction due to the history.
The LBR reports the mispredictions of indirect branches such that we
do not need to rely on elapsed cycles to attack indirect branches.

\PP{Attack procedure}
\autoref{f:indi} shows a detailed procedure of indirect branch
shadowing.
As mentioned previously, we set the target of the shadowed indirect
branch target as its next instruction to observe whether a branch
target misprediction happens or not due to the branch history.
We first explain the case where an unconditional branch has been
executed (Case 5).
\BC{1} An indirect branch of the victim code inside an enclave is
executed (\ie, taken) and the corresponding information is stored into
the BTB.
\BC{2} The enclave execution is interrupted and OS takes control.
\BC{3} The OS kernel enables the LBR and then executes the shadow code.
\BC{4} The BPU mispredicts the branch target of the shadowed
indirect branch due to the mismatched branch history.
\BC{5} The execution is rolled back and the shadow code jumps into the
correct target.
%
%
\BC{6} Finally, by disabling and retrieving the LBR, we can know that
the shadow code's indirect branch has been incorrectly predicted.

Next, we explain the case where an indirect branch has not been taken
(Case 6).
\BC{1} The enclave does not execute the indirect branch of the victim
code, so that the BTB does not have any information of the branch.
\BC{2} The enclave execution is interrupted and an OS takes control.
\BC{3} The OS kernel enables the LBR and then executes the shadow code.
\BC{4} The BPU correctly predicts the indirect branch's target
because there is no branch history.
%
%
\BC{5} Finally, by disabling and retrieving the LBR, we can know that the
shadow code's indirect branch has been correctly predicted,
implying that the victim code's indirect branch
has not been executed.

\PP{Inferring branch targets}
Unlike conditional and unconditional branches, an indirect branch can
have multiple targets such that just knowing whether it has been
executed or not would be insufficient to know the victim code's
execution.
Since the indirect branch is mostly used for representing a
\cc{switch-case} statement, it is also related to a number of
unconditional branches (\ie, \cc{break}) as an \cc{if-else} statement
does.
This implies that an attacker can identify which case block has been
executed by probing the corresponding unconditional branch.
Also, if an attacker can repeatedly execute a victim enclave
program with the same input, he or she can test the same indirect
branch multiple times while changing candidate target addresses to
eventually know the real target address by observing a correct branch
target prediction.

\begin{table}[!t]
  \centering
  \scriptsize
  \begin{tabular}{@{}rrcccc@{}}
\toprule
  \multirow{2}{*}{\bf Branch} & \multirow{2}{*}{\bf State}
    & \multirow{2}{*}{\textbf{BTB}} 
    & \multicolumn{2}{c}{\bf LBR} 
    & \multirow{2}{*}{{\bf Inferred}} \\
\cmidrule{4-5}  
    && 
    & {\bf Pred.} 
    & {\bf Elapsed Cycl.} & \\
\midrule
  \multirow{2}{*}{Cond.} & Taken        & \V & \V& -  & \V \\
                         & Not-taken    & -  & \V& -  & \V \\
\midrule
  \multirow{2}{*}{Uncond.} & Exec.      & \V & - & \V & \V \\
                           & Not-exec.  & -  & - & \V & \V \\
\midrule
  \multirow{2}{*}{Indirect} & Exec.     & \V & \V & - & \V \\
                            & Not-exec. & -  & \V & - & \V \\
  \bottomrule
\end{tabular}

  \caption{Branch types and states the branch shadowing attack can
    infer by using the information of BTB and/or LBR.}
  \label{tbl:attacks}
\end{table}

\autoref{tbl:attacks} summarizes the branch types and states our
branch shadowing attack can infer and the necessary information.

\subsection{Frequent Interrupt and Probe}
\label{subs:timer}
The branch shadowing attack needs to consider cases that change (or
even remove) BTB entries because they make the attack miss some branch
histories.
First, the size of the BTB is limited such that a BTB entry could be
overwritten by another branch instruction.
We empirically identified that the Skylake's BTB has 4,096 entries
where the number of ways is four and the number of sets is 1,024 (\autoref{subs:btb-clear}).
Due to its well-designed index hashing algorithm, we observed that
conflicts between two branch instructions located at different
addresses rarely happened.
But, no matter how, if more than 4,096 different branch instructions
have been taken, the BTB will highly likely be overflowed and we lose
some branch histories.
Second, a BTB entry for a conditional or an indirect branch can be
removed or changed due to a loop or re-execution of the same function.
For example, a conditional branch has been taken at its first run and
has not been taken at its second run due to the changes of the given
condition, removing the corresponding BTB entry.
A target of an indirect branch can also be changed according to
conditions, which change the corresponding BTB entry.
If the branch shadowing attack could not check a BTB entry before it
has been changed, it will lose the information.

To overcome this challenge, we interrupt an enclave process as
frequently as possible and check the branch history, by manipulating
the local APIC timer and the CPU cache.

\PP{Manipulating local APIC timer}
We manipulate the frequency of the local APIC timer in a recent
version of Linux that uses the TSC-deadline mode timer.
\begin{figure}[t!]
  \centering
  \input{code/apic.c.tex}
  \coderule
  \caption{Modified local APIC timer code of Linux kernel 4.4.23.  We
    changed \cc{lapic_next_deadline()} to manipulate the next TSC
    deadline and \cc{local_apic_timer_interrupt()} to launch the
    branch shadowing attack. We wrote a kernel module to change the
    exported global variables and function.}
  \label{c:apic}
\end{figure}
\autoref{c:apic} shows how we modified the \cc{lapic_next_deadline()}
function specifying the next TSC deadline and the
\cc{local_apic_timer_interrupt()} function called whenever a timer
interrupt is fired.
We made and exported two global variables and function pointers to
manipulate the behaviors of \cc{lapic_next_deadline()} and
\cc{local_apic_timer_interrupt()} with a kernel module:
\cc{lapic_next_deadline_delta} to change the delta;
\cc{lapic_target_cpu} to specify a virtual CPU running a victim
enclave process (via a CPU affinity); and
\cc{timer_interrupt_hook} to
specify a function to be called whenever a timer interrupt is
generated.
In our evaluation environment having an Intel Core i7 6700K CPU
(4GHz), we were able to have 1,000 as the minimum delta value; \ie, it
fires a timer interrupt about every 1,000 cycles.
Note that, in our environment, a delta value lower than 1,000 made the
entire system freeze because a timer interrupt was generated before an
old timer interrupt was handled by the interrupt handler.

We also counted how many CPU instructions can be executed between such
frequent timer interrupts by running a loop with an \cc{ADD}
instruction.
On average, about 48.76 \cc{ADD} instructions were executed between two
timer interrupts (standard deviation: 2.75)\footnote{The number of iterations was 10,000.
  We disabled Hyper-Threading, SpeedStep, TurboBoost, and C-States to reduce noise.}.
%
%
This implies that, by using this frequent timer, we can apply the
branch shadowing attack to a victim enclave process every 50th
instructions.

\PP{Disabling cache}
If we want to attack a very short loop having branches (\ie, shorter
than 50 instructions), the frequent timer interrupt would not be
enough.
To interrupt an enclave process more frequently, we selectively
disable the L1 and L2 cache of a CPU core running the victim enclave
process, by setting the cache disable (CD) bit of the CR0 control
register through a kernel module.
With the frequent timer interrupt and disabled cache, about 4.71
\cc{ADD} instructions were executed between two timer interrupts on
average (standard deviation: 1.96 with 10,000 iterations).
This would be enough to attack most branches.
One limitation of cache disabling is that it significantly slows
the execution of a victim enclave process such that the process may
notice it is under an attack.
Therefore, an attacker needs to carefully disable the cache only for
certain cases (\eg, when he or she recognizes a victim enclave process
is executing a function containing a very short loop).

\subsection{Virtual Address Manipulation}
\label{subs:memory}
To perform the branch shadowing attack, an attacker has to manipulate
the virtual addresses of a victim enclave process.
Since the attacker has already compromised an OS, manipulating the
page table to change virtual addresses is an easy task.
For simplicity, we assume the attacker disables the user-space
ASLR and modifies the Intel SGX driver for Linux (\cc{vm_mmap}) to
change the base address of an enclave, as shown in
\autoref{c:mal-sgx-driver}.
Also, the attacker puts an arbitrary number of \cc{NOP} instructions
before the shadow code to satisfy the alignment.
\begin{figure}[t!]
  \centering
  \input{code/mal-sgx-driver.c.tex}
  \coderule
  \caption{Modified Intel SGX driver to manipulate the base address of
    an enclave}
  \label{c:mal-sgx-driver}
\end{figure}

\subsection{Attack Synchronization}
\label{subs:synchronization}
Although the branch shadowing attack probes multiple branches in
each iteration, it is insufficient when a victim enclave
program is large.
An approach to overcome this limitation is to apply the
branch shadowing attack in a function level.
Namely, an attacker first infers functions a victim enclave program
either has executed or is currently executing and then probes branches
belonging to the functions.
If those functions contain entry points that can be invoked from
outside (via the \cc{EENTER} instruction) or rely on external calls, the
attacker can correctly identify them because they are controllable and
observable by the OS.
However, the attacker needs another strategy to infer the execution of
non-exported functions.

To find such executed functions, an attacker can create special shadow
code consisting of always reachable branches of target functions (\eg,
a conditional or unconditional branch located at the prologue).
By periodically executing the special shadow code, the attacker can
know which function has been executed and will run certain shadow code
for the function.

Also, we can use the page-fault side channel~\cite{xu:cca} to
synchronize attacks in terms of pages.
Since this channel allows an attacker to know a code page that is
about to be executed, he or she only needs to check functions located
in the code page.
But, this approach would not work when
a victim enclave process is secured with recent
studies~\cite{shinde:pigeonhole,costan:sanctum,tsgx} that prevent
page-fault side channels.
\section{Case Studies}
\label{s:study}
In this section, we explain how we can use the branch shadowing to
attack SGX applications and recent studies of securing SGX against the
controlled-channel attack.
\subsection{Attacking Enclave Applications}
We explain how the branch shadowing attack infers
fine-grained control-flow information of target SGX programs.
Specifically, we focus on examples in which the controlled-channel attack
cannot extract any information, \eg, control flows within
a single page.

\begin{figure}[t!]
  \centering
\begin{subfigure}[t]{\columnwidth}
  \centering
  \input{code/strtol.c.tex}
  \coderule
  \caption{Simplified \cc{strtol()}. The branch
    shadowing attack can infer the sign and length of an input
    number.}
  \label{c:strtol}
\end{subfigure}
~
\begin{subfigure}[t]{\columnwidth}
  \centering
  \input{code/vfprintf.c.tex}
  \coderule
  \caption{Simplified \cc{vfprintf()}. The branch shadowing attack can
    infer the format string and variable arguments.}
  \label{c:vfprintf}
\end{subfigure}
\caption{libc functions of Linux SGX SDK attacked by branch shadowing}
\end{figure}

\PP{Linux SGX SDK}
We attacked two libc functions, \cc{strtol()} and \cc{vfprint()},
supported by Linux SGX SDK.
\autoref{c:strtol} is a simplified \cc{strtol()} function that
converts a string into an integer.
%
%
By using the branch shadowing attack, we were able to infer the sign
of an input number by checking the branches in Lines 10--15.
Also, we could infer the length of an input number by checking the
loop branch in Lines 18--27.
In addition, when an input number was hexadecimal, we were able to
use the branch at Line 20 to know whether each digit was larger than nine.

\autoref{c:vfprintf} is a simplified \cc{vfprintf()} function used to print
a formatted string.
The branch shadowing attack was able to infer the format string by
checking the \cc{switch-case} statement in Lines 4--13 and the types
of input arguments to this function according the \cc{switch-case}
statement in Lines 15--23.
In contrast,
the controlled-channel attack cannot infer this information because
the functions called by \cc{vfprint()},
including \cc{ADDSARG()} and \cc{va_arg()}, are inline
functions.
No page fault sequence will be observed.

\begin{figure}[t!]
  \centering
  \input{code/mpi-montmul.c.tex}
  \coderule
  \caption{Montgomery multiplication (\cc{mpi_montmul()}) of mbed TLS.
    The branch shadowing attack can infer whether a dummy subtraction
    has performed or not.}
  \label{c:montmul}
\end{figure}

\PP{mbed TLS}
mbed TLS is a lightweight implementation of TLS.
We ported it to Intel SGX and tried to attack its RSA implementation,
which was not supported by Intel SGX SDK.
mbed TLS's RSA uses the Montgomery multiplication, as shown in
\autoref{c:montmul}, which has a dummy subtraction (Lines 24--27) to
prevent the well-known remote timing attack~\cite{remote-timing}.
The branch shadowing attack was able to detect the execution of this
dummy branch.
However, the controlled-channel cannot know whether a dummy
subtraction has happened because both real and dummy branches execute
the same function: \cc{mpi_sub_hlp()}.

\begin{figure}[t!]
  \centering
  \input{code/k-function.c.tex}
  \coderule
  \caption{Kernel function of LIBSVM. The branch shadowing attack can
    infer the kernel type.}
  \label{c:k-function}
\end{figure}

\PP{LIBSVM}
LIBSVM is a popular library supporting support vector machine (SVM)
classifiers.
We ported a classification logic of LIBSVM to Intel SGX because it
would be a good example of machine learning as a
service~\cite{ohrimenko:sgx-ml} while hiding the detailed parameters.
\autoref{c:k-function} shows the LIBSVM's kernel function code running
inside an enclave.
The branch shadowing attack can recognize the kernel type such as
linear, polynomial, and radial basis function (RBF) due to the
\cc{switch-case} statement in Lines 4--28.
Also, when a victim used an RBF kernel, we were able to infer the
number of features (\ie, the length of a vector) he or she used (Lines
11--20).

\begin{figure}[t!]
  \centering
  \input{code/lookup-builtin-method.c.tex}
  \coderule
  \caption{HTTP method lookup function in Apache's http module. The
    branch shadowing attack can infer the type of http method sent by
    clients.}
  \label{c:apache-lookup}
\end{figure}

\PP{Apache}
Apache is the most widely used web server.
We ported Apache by decoupling the original Apache program
such that some modules, such as the HTTP module, are secured by Intel SGX.
\autoref{c:apache-lookup} shows a lookup function of Apache to parse
the method of an HTTP request.
Due to its \cc{switch-case} statements, we can easily identify the
method of a target HTTP request, such as \cc{GET}, \cc{POST},
\cc{DELETE}, and \cc{PATCH}.
Since this function invokes either no function or \cc{memcmp()}, the
controlled-channel attack has no chance to identify the method.
%

\subsection{Attacking Side Channel Mitigations}
In this section, we review recent
studies~\cite{shinde:pigeonhole,costan:sanctum,tsgx} to secure Intel
SGX against page-fault and/or cache-timing attacks, and explain how
the branch shadowing attack can defeat them.
We also discuss how we can use the branch shadowing attack to break an
ASLR implementation in SGX~\cite{sgx-shield}, though it is outside the
scope of our threat model.

\begin{figure}[!t]
  \centering
  \begin{subfigure}[t]{\columnwidth}
    \centering
    \includegraphics[width=.5\textwidth]{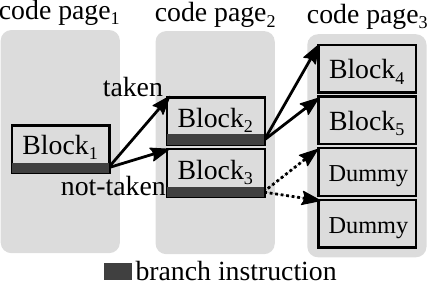}
    \caption{Deterministic multiplexing}
    \label{f:case-dm}
  \end{subfigure}
  \\
  \begin{subfigure}[t]{.49\columnwidth}
    \centering
    \includegraphics[width=.65\textwidth]{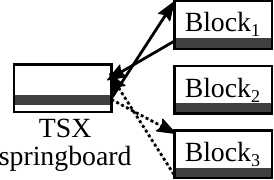}    
    \caption{T-SGX}
    \label{f:case-tsgx}
  \end{subfigure}
  \begin{subfigure}[t]{.49\columnwidth}
    \centering
    \includegraphics[width=.35\textwidth]{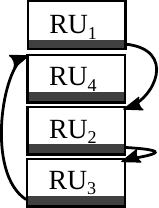}    
    \caption{SGX-Shield}
    \label{f:case-shield}
  \end{subfigure}
  \caption{Software-based methods to secure SGX. An attacker knows or
    can estimate the locations of branch instructions such that branch
    shadowing attacks are possible.}
  \label{f:case-papers}
\end{figure}

\PP{Deterministic multiplexing}
To prevent the page-fault side channel,
Shinde~\etal~\cite{shinde:pigeonhole} propose a deterministic
multiplexing technique to make all page accesses oblivious.
This technique is a weak form of the oblivious RAM (ORAM)
technique~\cite{path-oram,liu:ghostrider,mass:phantom,rane:raccoon},
but much faster than when developer-assisted compiler optimization
is applied (at most 1.29\X overhead).
The deterministic multiplexing works as follows.
First, it makes the execution tree of each function balanced by
introducing dummy (or decoy) branches and basic blocks
(\autoref{f:case-dm}).
This balanced execution tree is necessary to hide the behavior and
execution time of a function because it can reveal which basic blocks
of the function have been executed.
Next, the deterministic multiplexing puts all real and dummy code
blocks at the same execution level into the same code page and all
data blocks that the code blocks will access into the same data page.
This ensures that whether an enclave process is executing a real or dummy
block, a page fault will occur at the same page.
Thus, monitoring page fault sequences no longer reveals the control
flows of a victim enclave process.

However, the branch shadowing attack can easily defeat the
deterministic multiplexing technique because this attack can know
whether a victim enclave process is executing a real or dummy block by
using the branch history, not the page faults.
That is, the selective execution of real or dummy branches according to
condition evaluation cannot hide any secrets from the branch shadowing
attack.
One possible way to improve the deterministic multiplexing technique
is to always to execute both real and dummy branches as
Raccoon~\cite{rane:raccoon} does.
However, it is difficult to use in practice due to huge
performance overhead (21.8\X).

\PP{T-SGX}
Shih~\etal~\cite{tsgx} propose T-SGX, which is a software-based technique
to detect a page fault in user space by using an existing Intel CPU
instruction: Transactional Synchronization Extensions (TSX).
Intel TSX allows a user-level process to know whether a memory access
generates a page fault or not before it is delivered to an OS.
Therefore, with Intel TSX, an enclave process can detect suspicious
page faults and terminate its execution, whose effects would be the
same as proposals demanding hardware
modifications~\cite{costan:sanctum,shinde:pigeonhole}.

T-SGX protects each basic block by individually wrapping it with Intel
TSX and makes each of them jump to each other through a springboard
page to enforce control flows (\autoref{f:case-tsgx}).
However, the branch shadowing attack can easily recognize which blocks
have been executed by probing those branch instructions, implying that
T-SGX cannot be used to detect or prevent the branch shadowing attack.

\PP{SGX-Shield}
Seo~\etal~\cite{sgx-shield} develop SGX-Shield, which is an enclave
program to load the code consisting of randomization units (RUs) while
randomizing their locations in place, \ie, it implements fine-grained
ASLR (\autoref{f:case-shield}).
Since a malicious OS is no longer able to know the exact addresses of
the target branch instructions due to randomization, it is difficult
to directly apply the branch shadowing attack.

However, an attacker can infer the execution sequence of RUs because of
the following limitations of SGX-Shield.
First, SGX-Shield does not support live re-randomization such that the
locations of all branch instructions are not changed during its
execution.
Second, the sizes of RUs are fixed (32 or 64 bytes) and their
addresses are aligned to avoid any decoding errors.
Since the last instruction of an RU is always a branch instruction to
jump into the next RU, an attacker can identify whether it has been
executed by testing the branch instruction.
By repeating it against all RUs, the attacker will obtain the
execution sequence eventually.

\PP{Sanctum}
Costan~\etal~\cite{costan:sanctum} design a new hardware-based TEE,
called Sanctum, which is a secured version of Intel SGX, built on top
of the RISC-V~\cite{risc-v,risc-v2} Rocket Core.
Sanctum's goals are detecting page-fault-based attacks and preventing
cache-timing attacks.
First, to detect page-fault-base attacks, Sanctum lets an enclave
process know whether a page fault is occurring without the help
of an OS (Shinde~\etal~\cite{shinde:pigeonhole} also mention a similar
hardware design.)
The enclave process then inspects whether the page fault is legitimate
 and terminates its execution when there is a security problem.
Second, to prevent cache-timing attacks, Sanctum implements a
page-coloring technique to partition the last-level cache (LLC).
In Sanctum, physical addresses are shifted before being stored in the LLC,
so that an OS cannot know the cache set storing its target memory
page.

However, since the branch shadowing attack neither generates any page
faults nor relies on physical addresses, such countermeasures are
irrelevant to this attack.
Further, Sanctum aims to bring minimal modifications to the RISC-V
Rocket Core, which also supports static and dynamic branch prediction.
This implies that, by manipulating virtual addresses, we
can perform branch shadowing attacks against the Sanctum's enclave
unless Sanctum obfuscates branch prediction behaviors.

\section{Countermeasures}
\label{s:counter}
In this section, we introduce our hardware-based and
software-based countermeasures against the branch shadowing attack.

\subsection{Hardware-based Countermeasure}
\label{subs:btb-clear}

\begin{figure*}[!t]
\centering
\input{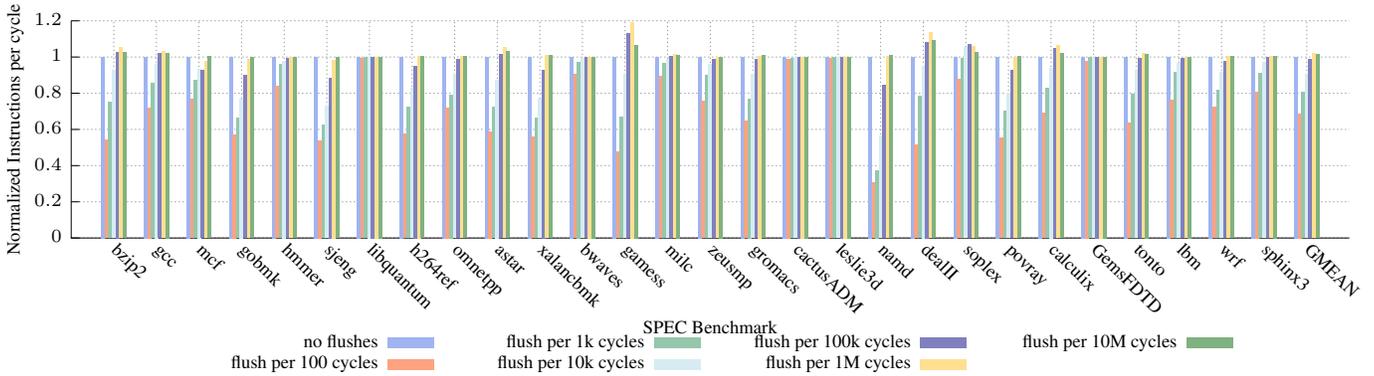}
\caption{Instructions per cycle of SPEC benchmark in terms of frequency of BTB + BPU flushing.}
  \label{fig:spec-btb-ipc}
\end{figure*}

\begin{figure}[!t]
\centering
\input{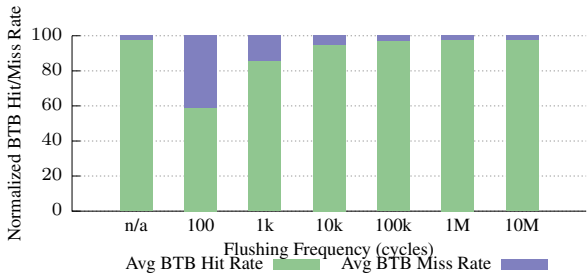}
\caption{Average BTB hit/miss rate for SPEC06 w.r.t. frequency of BTB + BPU flushing.}
\label{fig:btb-hit-rate}
\end{figure}

\begin{figure}[!t]
\centering
\input{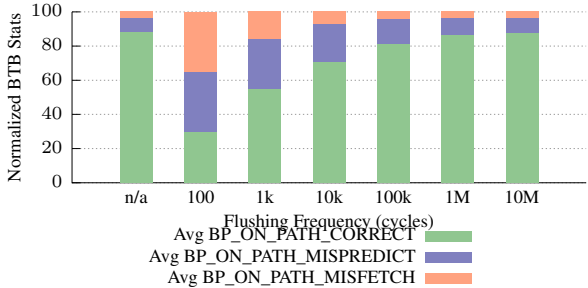}
\caption{Average BTB stats for SPEC06 w.r.t. frequency of BTB + BPU flushing.}
\label{fig:spec-btb-stat}
\end{figure}

The micro-architectural state of branch execution is maintained in two important
structures: BTB and BPU.
These are not necessarily monolithic structures, and they may be further
divided into sub-structures depending on the implementation. For instance, the
BTB may comprise of a different unit for indirect branches known as an iBTB.
These structures are implemented per hardware core, and on systems that use
Simultaneous Multi-Threading (SMT), they are generally shared between all the
hardware threads of the core. On modern Intel processors with hyperthreading, we
confirmed that the BTB state is shared between different SMT threads (hyperthreads) by
creating set conflicts in the BTB between two different hyper threads.
To mitigate the class of security vulnerabilities described in this paper, we
need to prevent two sources of information leakage:
hyperthreads running in the same core;
and the kernel and user code on the same hardware thread.
Preventing leakage between hyperthreads is only possible if different hardware
threads use different structures, or if hyperthreading is disabled (\autoref{subs:advanced}).
In order to
prevent the leakage of information on the same hardware thread, we need to
ensure that all branch related states are flushed whenever the context switches
to or from enclave mode. Whenever an enclave context switch (via the
\cc{EENTER}, \cc{EEXIT}, or \cc{ERESUME} instruction or AEX) happens,
we need to flush the BTB and BPU state. Since the BTB and BPU benefit from
local and global branch execution history, there would be a performance penalty
if these structures are flushed too frequently.

\begin{table}[!t]
  \centering
  \scriptsize
  \begin{tabular}{@{}ll@{}}
  \toprule
  \textbf{Parameter} & \textbf{Value} \\
  \midrule
  CPU     & 4 GHz out of order core, 4 issue width, 256 entry ROB \\
  L1 cache & 8 way 32 KB I-cache + 8 way 32 KB D-cache \\
  L2 cache & 8 way 128 KB \\ 
  L3 cache & 32 way 8 MB \\
  BTB      & 4 way 1,024 sets \\ 
  BPU      & gshare, branch history length 16 \\
  \bottomrule
\end{tabular}

  \caption{MacSim Simulation parameters}
  \label{macsimparams}
\end{table}

We aim to determine the
performance impact of flushing these structures at different frequencies in a
cycle level out-of-order microarchitecture simulator,
MacSim~\cite{kim2012macsim}.
The details of our simulation parameters are
listed in \autoref{macsimparams}. The BTB is modeled after the BTB in Intel Skylake
processors. We used a method similar
to~\cite{uzelac2009experiment,TheBTBin80:online} to reverse engineer the BTB
parameters. From our experiments, we found that the BTB is organized as a 4-way
set associative structure with a total of 4,096 entries.  We model a simple
branch predictor, gshare~\cite{McFarling93}, for the simulation. Current Intel
processors use more advanced predictors, but the specifics are not very
important for these experiments. We use 200 million instruction long traces
from the SPEC06 benchmark suite for simulation and flush the BTB and BPU
periodically at varying frequencies.

\autoref{fig:spec-btb-ipc} shows the normalized instructions per cycle (IPC)
for different flush frequencies. We found that if the flush frequency is higher
than 100K cycles, it has a negligible impact on the performance. At a flush
frequency of 100K cycles, the performance impact is lower than 2\% and at 1
million cycles, it is negligible. \autoref{fig:btb-hit-rate} shows the BTB hit
rate, whereas \autoref{fig:spec-btb-stat} shows the BPU \textit{correct},
\textit{incorrect} (direction prediction is wrong), and \textit{misfetch}
(target prediction is wrong) percentages. The BTB and BPU statistics are also
virtually indistinguishable beyond a flush frequency of 100K cycles.

In a 4GHz CPU, if we assume that
the interval between interrupts (or AEX) is 100K cycles,
there would be 10,000 interrupts per second.
According to our measurements, about 250 and 1,000 timer interrupts
are generated per second in Linux (version 4.4) and Windows 10,
respectively.
Thus, if there is no I/O device generating too many interrupts, the
flush frequency of 100K cycles would be reasonable.

\subsection{Software-based Countermeasure}
The hardware-based countermeasure can effectively prevent the branch
shadowing attack, but we cannot be sure when and whether such hardware
changes can be realized.
Especially, if such changes cannot be done with micro code updates,
Intel CPUs already deployed in the markets would have no
countermeasure against the branch shadowing attack.

Possible software-based countermeasures against the branch shadowing
attack are to
remove branches~\cite{ohrimenko:sgx-ml} or to use the
state-of-the-art ORAM technique, Raccoon~\cite{rane:raccoon}.
Ohrimenko~\etal~\cite{ohrimenko:sgx-ml}'s data-oblivious machine
learning algorithms try to eliminate all branches by using a
conditional move instruction, \cc{CMOV}.
However, their approach is algorithm-specific, \ie, we cannot apply it
to general applications.
%
%
Raccoon~\cite{rane:raccoon} always executes both paths of a
conditional branch, such that it can hide whether the branch has been
really taken from a branch shadowing attack.
%
%
But, its performance overhead is high (21.8\X).

\PP{Zigzagger}
We propose a practical, compiler-based mitigation scheme
against the branch shadowing attack, 
called \emph{Zigzagger}.
The basic idea of Zigzagger is to obfuscate
a set of branch instructions
into a single indirect jump.
However, it is not straightforward
to compute the target block of each branch
without relying on conditional jumps because
conditional expressions would become very complex when
we need to handle nested branches.
%
In Zigzagger,
we solved this problem by utilizing
a \cc{CMOV} instruction,
which performs a conditional \cc{MOV} operation,
and introducing
a sequence of non-conditional jump instructions
in lieu of each branch.
Zigzagger's approach has several benefits:
1) in terms of security, it provides a first line of
protection on each branch blocks
and explodes the potential flows
in an enclave program;
2) in terms of performance,
the unconditional jumps are much more favorable
to instruction pipelining;
3) in terms of practicality,
Zigzagger's transformation does not require complex
analysis of code semantics
(\ie, possible to implement it as a compiler pass).
Furthermore,
Zigzagger's execution pattern---%
back-and-forth jumps between the converted branch set and the
Zigzagger's trampoline---%
practically increases the bar for de-obfuscating
the fine-grained control-flow of the protected enclave problem.
It is worth noting that
this countermeasure is not specific to Intel SGX nor
the branch shadowing attack proposed in this paper;
we can use this approach to mitigate other
types of branch-based timing attacks.

\begin{figure}[t!]
  \centering
  \begin{subfigure}[t]{\columnwidth}
    \centering
    \includegraphics[width=.6\textwidth]{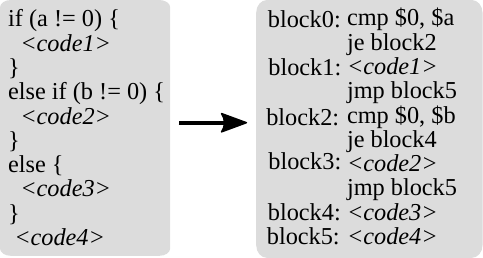}
    \caption{An example code snippet.
      It selectively executes a branch block
      according to \cc{a} and \cc{b} variables.}
    \label{f:trampoline-orig}
  \end{subfigure}
  ~
  \begin{subfigure}[t]{\columnwidth}
    \centering
    \includegraphics[width=.7\textwidth]{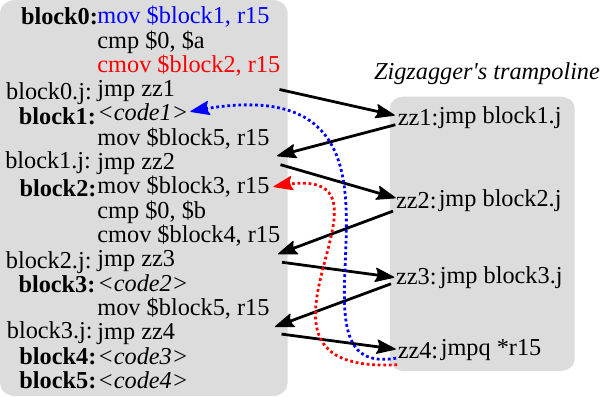}
    \caption{The protected code snippet by Zigzagger.
      All branch instructions are executed
      regardless of \cc{a} and \cc{b} variables.
      An indirect branch in the trampoline
      and \cc{CMOV} instructions in the translated code
      are used to obfuscate
      the final target address. 
      Note that \cc{r15} is reserved in Zigzagger
      to store the target address.}
    \label{f:trampoline-trans}
  \end{subfigure}
  \caption{Securing an example code snippet with Zigzagger.}
  \label{f:trampoline}
\end{figure}

\autoref{f:trampoline} shows how Zigzagger transforms
an example code snippet
having \cc{if}, \cc{else-if}, and \cc{else} blocks.
It converts all conditional and unconditional branches into
unconditional branches targeting Zigzagger's trampoline
that jumps
back-and-forth with the converted branches and finally jumps into the
real target address in a reserved register \cc{r15} stored before
jumping into the Zigzagger.
It reserves the register for performance reasons;
for programs that can utilize more registers, 
it can potentially use the main memory instead,
but reserving \cc{r15} in SGX has negligible performance overhead%
~\cite{tsgx}.
To emulate conditional execution without using conditional jump,
we use \cc{CMOV} instructions: 
\eg, the \cc{CMOV} instructions in
\autoref{f:trampoline-trans} update \cc{r15} only when \cc{a} or \cc{b} is zero.
Otherwise, these instructions are treated as \cc{NOP} instructions.
Since all of the unconditional branches are executed almost
simultaneously in sequence,
an attacker has difficulty recognizing the current instruction pointer;
our APIC timer trick is not fine-grained enough
to distinguish each branches in practice (\autoref{subs:timer}).
At last,
the indirect branch in Zigzagger's trampoline
now has five different target addresses,
obfuscating potential target addresses.

\PP{Implementation}
We implemented Zigzagger in LLVM 4.0
as an LLVM pass
that converts branches in each function
and constructs the required trampoline.
We also modified the LLVM backend
to reserve the \cc{r15} register.
We observed that when a function has many branches,
making them share a single trampoline in Zigzagger
introduces non-negligible performance overhead
due to frequent jumps.
To avoid this problem,
our implementation provides a knob to configure
the number of branches
that each trampoline manages and randomly assigns
branches to each trampoline.
Note that such merging-based optimization
trades the security for performance,
but we believe it becomes more useful in practice
(\eg, selectively applying to security-sensitive routines).

\begin{table}[!t]
  \centering
  \scriptsize
  \begin{tabular}{@{}lrr@{~~~}r@{~~~}r@{~~~}r@{~~~}r@{}}
  \toprule
  \textbf{Benchmark} & {\bf Baseline} & \multicolumn{5}{c}{\bf Zigzagger} \\
                     & {\bf (iter/s)} & \multicolumn{5}{c}{\bf \#Branches (overhead)} \\
  \cmidrule(){3-7}
                     &                & {\bf 2} & {\bf 3}  & {\bf 4} & {\bf 5} & {\bf All} \\
\midrule
 \cc{numeric sort}     & 967.25 &  1.05\X & 1.11\X & 1.12\X & 1.13\X & 1.15\X \\
 \cc{string sort}      & 682.31  & 1.08\X & 1.15\X & 1.18\X & 1.15\X & 1.27\X \\
 \cc{bitfield}         & 4.5E+08 & 1.03\X & 1.10\X & 1.14\X & 1.18\X & 1.31\X \\
 \cc{fp emulation}     & 96.204  & 1.10\X & 1.21\X & 1.15\X & 1.27\X & 1.35\X \\
 \cc{fourier}          & 54982  &  0.99\X & 0.99\X & 1.01\X & 1.01\X & 1.01\X \\
 \cc{assignment}       & 35.73  &  1.36\X & 1.56\X & 1.50\X & 1.55\X & 1.90\X \\
 \cc{idea}             & 10,378  & 2.16\X & 2.16\X & 2.18\X & 2.19\X & 2.19\X \\
 \cc{huffman}          & 2478.1  & 1.59\X & 1.46\X & 1.61\X & 1.63\X & 1.81\X \\
 \cc{neural net}       & 16.554  & 0.75\X & 0.77\X & 0.85\X & 0.86\X & 0.89\X \\
 \cc{lu decomposition} & 1,130   & 1.04\X & 1.09\X & 1.08\X & 1.11\X & 1.17\X \\
\midrule
GEOMEAN                &         & 1.17\X & 1.22\X & 1.24\X & 1.26\X & 1.34\X \\
\bottomrule
\end{tabular}

  \caption{Overhead of the Zigzagger approach according to the number
    of branches belonging to each Zigzagger}
  \label{tbl:nbench}
\end{table}
Our proof-of-concept implementation of Zigzagger, which
provides full protection,
imposes 1.34\X performance overheads,
when evaluating it with the \cc{nbench} benchmark suite
(\autoref{tbl:nbench}).
With optimization (\ie, merging $\le 3$ branches
into a single trampoline),
the average overhead becomes less than 1.22\X.
Note that reserving a register in our microbenchmark
results in 4\%--50\% performance improvement.

\section{Discussion}
\label{s:discuss}
In this section, we explain some limitations of the branch shadowing
attack and discuss possible advanced attacks.

\subsection{Limitations}
An important limitation of the branch shadowing attack is that it
cannot distinguish a not-taken conditional branch from a not-executed
conditional branch because, in both cases, the BTB has no information about the branch;
the static branch prediction rule is applied.
Also, the branch history attack cannot distinguish an indirect branch
to the right next instruction from a not-executed indirect branch
because their predicted branch targets are the same.
Therefore, an attacker has to probe a number of correlated branches
(\eg, unconditional branches in \cc{else-if} or \cc{case} blocks) to
overcome this limitation.

\subsection{Advanced Attacks}
\label{subs:advanced}
We introduce two advanced attacks based on the branch shadowing attack:
hyperthreaded branch shadowing attack and blind branch shadowing
attack.

\PP{Hyperthreaded branch shadowing}
Since two hyperthreads simultaneously running in the same physical
core share the same BTB, a malicious hyperthread is able to attack a
victim enclave hyperthread by using BTB entry conflicts, if a
malicious OS gives the address information of the victim to it.
We found that branch instructions with the same low 16-bit
address were mapped into the same BTB set.
Thus, a malicious hyperthread can monitor a BTB set for evictions by
filling the BTB set with four branch instructions.
The BTB clearing (\autoref{subs:btb-clear}) cannot prevent this attack
because no enclave mode switch happens.
However, this attack cannot precisely identify the higher order bits of the
victim branch's address yet since they aren't used in set index calculation.   
We plan to reverse engineer the BTB's characteristics in more detail to
determine whether we can obtain the exact address of taken branches.

\PP{Blind branch shadowing}
A blind branch shadowing attack is an attempt to probe the entire or
selected memory region of a victim enclave process to detect any
unknown branch instructions.
This attack would be necessary if a victim enclave process has
self-modifying code or uses remote code loading, though it is outside the
scope of our threat model (\autoref{subs:model}).
In the case of unconditional branches, blind probing is easy and effective
because it does not need to infer target addresses.
However, in the case of conditional and indirect branches, blind probing
needs to consider branch instructions and their targets simultaneously
such that the search space would be huge.
We plan to consider an effective method to minimize the search space
to know whether this attack is practical.

\section{Related work}
\label{s:relwk}
This section introduces studies related with our work
including studies on SGX and microarchitectural side channels.
%

\PP{Intel SGX.}
The strong security guarantee provided by SGX has drawn significant
attention from the research community.
Several security applications of SGX have been proposed, including
secure data analysis~\cite{schuster:vc3,ohrimenko:sgx-ml},
secure distributed computing~\cite{dinh:m2r,securekeeper}, and
secure networking service~\cite{shih:snfv,secure-cbr}.
Also, researchers implemented library OSes for
SGX~\cite{baumann:haven,arnautov:scone} to
run existing applications inside an enclave
without any modifications.
The security properties of SGX itself are also being intensively
studied.
For example, Sinha~\etal~\cite{sinha:moat, sinha:confidential} develop
tools to verify the confidentiality of enclave programs.

However, the authors of the above-mentioned projects do not consider the
potential security attacks against SGX.
Xu~\etal~\cite{xu:cca} and Shinde~\etal~\cite{shinde:pigeonhole}
demonstrate the first side-channel attack on SGX by leveraging the
fact that SGX relies on OS for memory resource management.
The attack is done by intentionally manipulating the page table to
trigger a page fault and using a page fault sequence to infer the secret
inside an enclave.
Weichbrodt~\etal~\cite{weichbrodt:asyncshock} also show
how a synchronous bug can be exploited to attack SGX applications.

To address the page-fault-based side-channel attack,
Shinde~\etal~\cite{shinde:pigeonhole} have proposed an ORAM-like
scheme that can effectively obfuscate the memory access pattern of
the enclave program, but it suffers from significant performance
overhead.
Shih~\etal~\cite{tsgx} have proposed a compiler-based solution using
Intel TSX to detect any suspicious page faults inside an enclave.
Also, Costan~\etal~\cite{costan:sanctum} have proposed a new enclave design
to prevent both page-fault and cache-timing side channels.
Finally, Seo~\etal~\cite{sgx-shield} have enforced fine-grained ASLR on
enclave programs, which can raise the bar of
exploiting any vulnerabilities and
inferring control flow with page-fault sequences.
However, we demonstrated that none of these solutions can mitigate the
branch shadowing attack.

\PP{Microarchitectural side channel}
Numerous researchers have considered the security problems of
microarchitectural side channels.
The most popular and well-studied microarchitectural side channel is a
CPU cache timing channel first developed by
\cite{kocher1996timing,kelsey1998side,page2002theoretical} to break
cryptosystems.
This attack is further extended to be conducted in the public cloud
setting to recognize co-residency of virtual
machines~\cite{ristenpart2009hey,zhang2011homealone}.
Recently, several researchers further improved this attack to exploit
the last level cache~\cite{ssa,llc-practical} and realize a low-noise
cache storage channel~\cite{cache-storage-channel}.
The CPU cache is not the sole source of the microarchitectural side
channel.
For example, Hund~\etal~\cite{hund:timing} exploits the translation
lookaside buffer (TLB) timing channel to break the kernel ASLR.
Also, researchers improve this attack by exploiting other side
channels in the Intel TSX~\cite{jang:drk}, a \cc{PREFETCH}
instruction~\cite{gruss:prefetch}, and the BTB~\cite{evtyushkin:btb},
respectively.
%
%
Note that Ge~\etal~\cite{ge:microarchitectural} publish a comprehensive
survey of microarchitectural side channels.

\section{Conclusion}
\label{s:conclusion}
A hardware-based TEE is a promising technology to realize the truly
secure public cloud, but, without serious security analysis, no one is
willing to trust and use the TEE.
Especially, a lack of thorough analysis of side channels is
problematic because it is difficult to ensure that a TEE is completely
free from any side channels.
In this paper, we explored a new side-channel attack against Intel
SGX, called a branch shadowing attack, which can precisely identify fine-grained
(basic-block-level) control flows executed inside an enclave.
%
%
We proposed a hardware-based countermeasure that clears the branch
history during enclave mode switch and a software-based
mitigation that makes branch executions oblivious.
%
%


\bibliographystyle{IEEEtranS}
\bibliography{p,sslab,tee,conf}
\end{document}